\documentclass[3p]{elsarticle}

\usepackage{a4wide}
\usepackage{float}
\usepackage{amsmath}
\usepackage{amssymb}
\usepackage{graphicx}
\usepackage[latin1]{inputenc} %Kann auch Umlaute etc. erkennen
\usepackage{rotating}
\usepackage{setspace}
\usepackage{subfigure}
\usepackage{sidecap}
\usepackage{upgreek}
\usepackage{microtype}
\usepackage{url}
\usepackage{lineno}
\begin{document}
\begin{frontmatter}

%\pagenumbering{Roman}

%------------------------------------ Title ------------------------
\title{Challenges for Silicon Pixel Sensors at the European XFEL}

\renewcommand{\thefootnote}{\fnsymbol{footnote}}

%\author{ J.~Schwandt$^{a,}$\thanks{Corresponding author.}~, E.~Fretwurst$^a$, R.~Klanner$^a$ and J.~Zhang$^a$\\
%\llap{$^a$}Institute for Experimental Physics, University of Hamburg\\
%  Luruper Chaussee 149, D-22761 Hamburg, Germany\\
 % E-mail: \email{joern.schwandt@desy.de}}

\author[]{Robert Klanner$^{a,}$ \corref{cor1}}
\author[]{Julian Becker$^b$}
\author[]{Eckhart Fretwurst$^a$}
\author[]{Ioana Pintilie$^c$}
\author[]{Thomas P\"ohlsen$^a$}
\author[]{Jörn~Schwandt$^a$}
\author[]{and Jiaguo~Zhang$^a$}
\cortext[cor1]{Corresponding author. Email address: Robert.Klanner@desy.de. Telephone: +49 40 8998 2558.}

 \address{$^a$~Institute for Experimental Physics, University of Hamburg, Hamburg, Germany}
%, Luruper Chaussee 149, 22761 Hamburg, Germany}
 \address{$^b$~DESY, Hamburg, Germany}
% , Notkestrasse 85, 22607 Hamburg, Germany}
 \address{$^c$~National Institute of Materials Physics, Bucharest, Romania}
% , Atomistilor Str.~105bis, PO Box MG7, Magurele, 077125 Bucharest, Romania}

 \begin{abstract}

 A systematic experimental study of the main challenges for silicon-pixel sensors at the European XFEL is presented.
 The high instantaneous density of X-rays and the high repetition rate of the XFEL pulses result in signal distortions due to the plasma effect and in severe radiation damage.
 The main parameters of X-ray-radiation damage have been determined and their impact on $p^+n$ sensors investigated.
 These studies form the basis of the optimized design of a pixel-sensor for experimentation at the European XFEL.

\end{abstract}

 \begin{keyword}
  XFEL \sep silicon-pixel sensor \sep plasma effect \sep charge losses \sep X-ray-radiation damage \sep sensor optimization.
 \end{keyword}

 \end{frontmatter}

\section{Introduction}
% A citation ------------------
%  \cite{AGIPD}

 The high instantaneous intensity and the high repetition rate of 4.5 MHz of the European X-Ray Free-Electron Laser (XFEL) \cite{XFEL, Tschentscher:2011} pose new challenges for imaging detectors.
 The specific requirements for the detectors include a dynamic range of 0, 1 to more than 10$^4$ photons of typically 12.4~keV per pixel for an XFEL pulse duration of less than 100 fs, and a radiation tolerance for doses up to 1~GGy for 3 years of operation~\cite{Graafsma:2009}.
 In addition, the sensors should have good detection efficiencies for X-rays with energies between ~3 and ~20 keV, and minimal inactive regions at their edges.
 Within the AGIPD Collaboration~\cite{Henrich:2011, Henrich1:2011, AGIPD} the Hamburg group has  studied the consequences of these requirements for $p^+n$-silicon sensors and an optimized design for the AGIPD sensor  been presented.

 High instantaneous X-ray intensities cause the so-called plasma effect~\cite{Tove:1967, Becker:2010}, which results in a significant change of the  signal shape and of the spatial distribution of the collected charges, compared to single-photon detection.
 The plasma effect has been studied using a $p^+n$ strip sensor read out by a multi-TCT system (Transient Current Technique) \cite{Becker:Thesis} for charge carriers generated by sub-nanosecond focused light with absorption lengths between 3.5~$\upmu$m and 1~mm (660 to 1060~nm wavelengths).
 Results on the pulse shape and point-spread function as function of charge-carrier density, and their impact on the choice of the resistivity of the silicon and the operating voltage will be presented.

 The high X-ray dose results in an increase of the oxide-charge density and in the formation of traps at the Si-SiO$_2$ interface.
 Detailed C/G-V and TDRC (Thermal Dielectric Relaxation Current) measurements on MOS capacitors from different vendors can be described by 3 dominant interface traps.
 Their locations in the silicon-band gap, and their capture cross sections and densities as function of X-ray dose~\cite{Zhang:2012, Zhang:2011} have been determined.
 I-V measurements on Gate-Controlled Diodes were performed to determine the surface-current density at the depleted Si-SiO$_2$ interface.
 In addition, the annealing behavior~\cite{Zhang1:2012} has been measured and described by models.
% The results of these studies will be presented.

 The X-ray-radiation damage has a major impact on the dark current, and the charge densities, and the electric fields in the vicinity of the Si-SiO$_2$ interface and  $p^+$ implants.
 They in turn, influence sensor properties, like breakdown and depletion voltage, inter-electrode capacitances and charge-collection efficiency.
 The latter have been studied on sensors before and after irradiation to 1 MGy (SiO$_2$) with ~12~keV X-rays using I-V and C-V measurements, and the multi-channel TCT using focused light of 3.5~$\upmu$m absorption length (660 nm wavelength) from a sub-nanosecond laser~\cite{Poehlsen:2012}.
 From these measurements the charge collection efficiencies and the extension of the accumulation layers at the Si-SiO$_2$ interface as function of dose, biasing history and ambient humidity have been determined.
 The experimental findings are explained with the help of detailed TCAD simulations~\cite{Poehlsen:2012}.
% A summary of these fairly complex results will be given.

 Finally, with the help of extensive TCAD simulations which use the results of above measurements, the AGIPD sensor ($512 \times 128$ $p^+$ pixels of 200~$\upmu$m $\times $~200~$\upmu$m on 500~$\upmu$m thick $n$-type silicon) has been designed for operation at radiation doses between 0 and 1~GGy~\cite{Schwandt:2012, Schwandt1:2012}.
% The strategy and the results of the optimization, and the expected sensor parameters like dark current, breakdown voltage and inter-pixel capacitance will be presented.

\section{Plasma effect}

 The plasma effect in solid state sensors, which is well known from the measurements of ions and nuclear fragments, occurs when the density of electron-hole ($eh$) pairs produced by the radiation is large, typically of the order or larger than the doping of the crystal.
 For 10$^5$ photons of 12.4~keV in a pixel of (200~$\upmu $m)$^2$ the density of $eh$~pairs is a few times $10^{13}$~cm$^{-3}$, compared to the typical doping of 10$^{12}$~cm$^{-3}$.
 On a time scale of picoseconds after their generation, holes and electrons move in opposite directions in the electric field and form a neutral $eh$ plasma with a field-free region in the plasma, surrounded by high-field regions.
 The plasma erodes by ambipolar diffusion, resulting in a delayed charge collection and a spreading of the charges by diffusion and electrostatic repulsion.
 For a simulation, including a movie accessible from the on-line version, we refer to~\cite{Gaertner:2010}.
 For the experimental studies~\cite{Becker:2010, Becker:Thesis} focussed sub-ns laser light of different wavelengths was used to create $eh$ pairs in a silicon-strip sensor with 80~$\upmu $m pitch and the transients on the strips were measured using fast amplifiers and a 2.5~GHz scope.
 The wavelengths of the lasers were 660 and 1015~nm. The corresponding absorption lengths in silicon are 3.5 and 250~$\upmu$m,  simulating X-rays of 1 and 12.4~keV, respectively.

 \begin{figure}
   \centering
	\includegraphics[width=7cm]{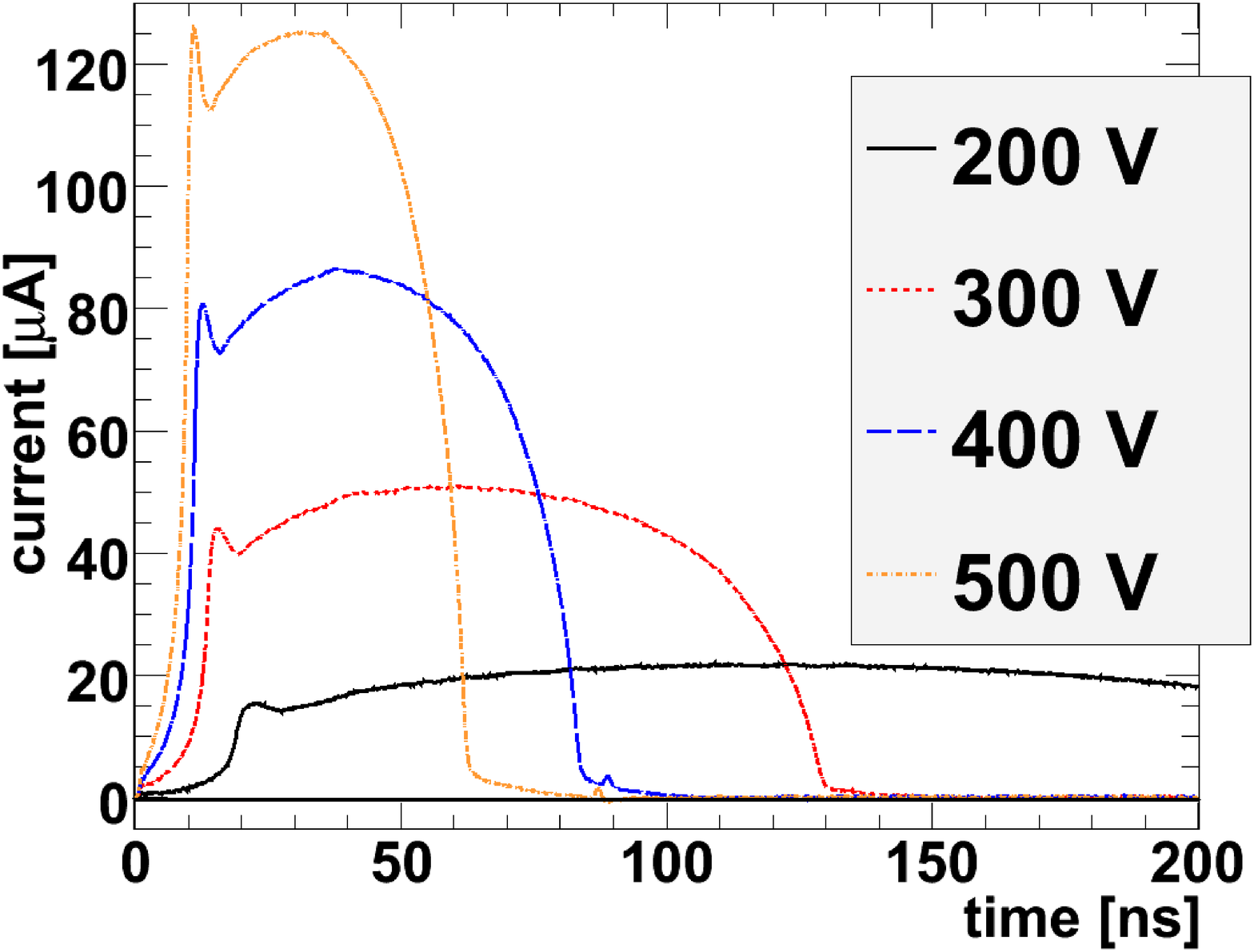}
	\includegraphics[width=7cm]{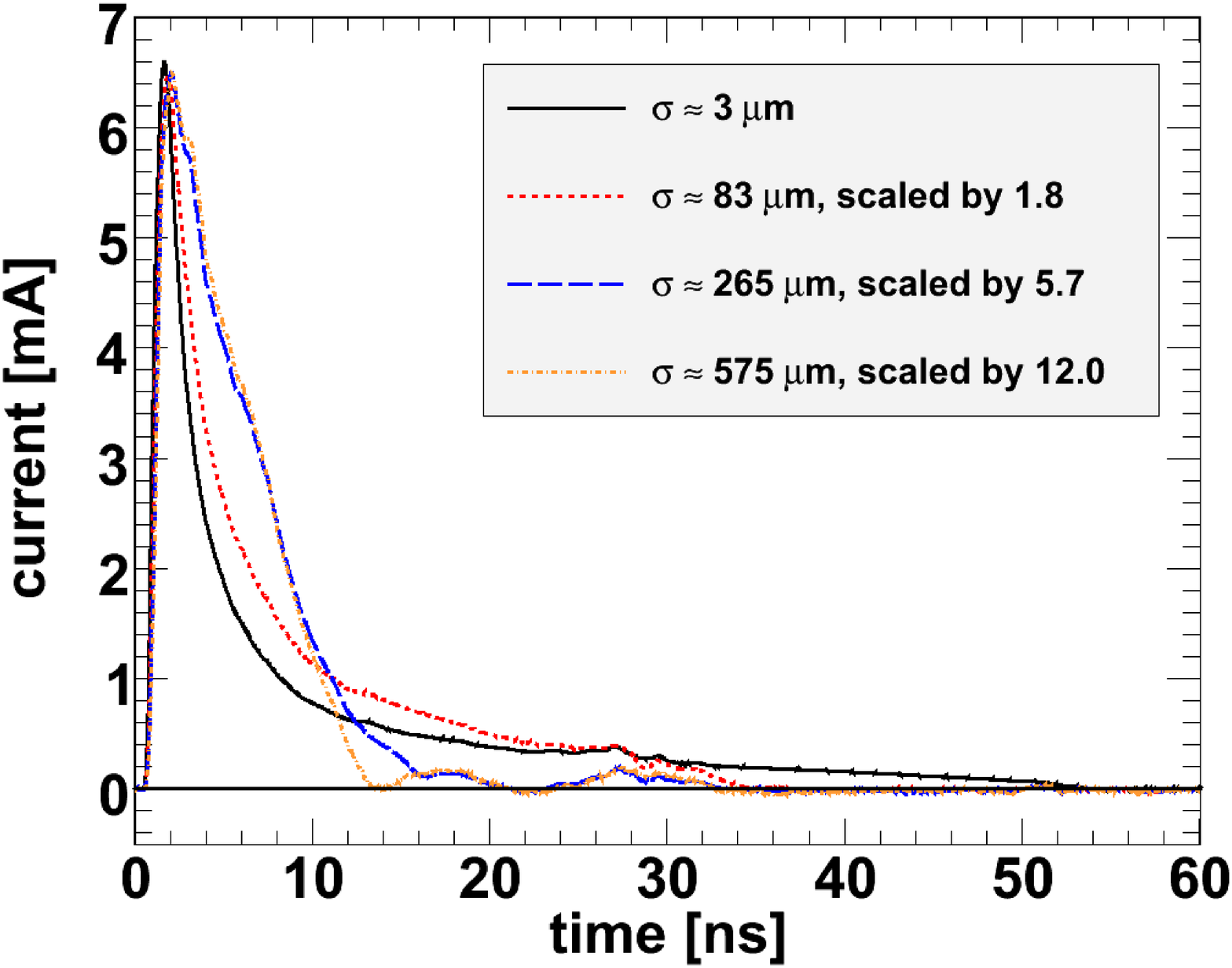}	
   \caption{Current transients for a 450~$\upmu$m silicon  sensor with a depletion voltage of 155~V.
   Left: Dependence on applied voltage for 660~nm laser light focussed to an $rms$ value of 3~$\upmu$m and an intensity corresponding to 3.2~$\times 10^{5}$ photons of 1~keV.
   Right: For a voltage of 500~V the dependence on the $rms$ focussing $\sigma $ of the 1015~nm laser light for an intensity  corresponding to 1.5~$\times 10^{5}$ photons of 12~keV. }
  \label{fig:PlasmaTime}
 \end{figure}

 The left side of figure~\ref{fig:PlasmaTime} shows the current transients for a 450~$\upmu $m thick sensor  for different applied voltages.
 The $eh$ pairs produced by the 660~nm laser corresponded to $3.2 \times 10^5$ photons of 1~keV focussed to an $rms$ value of 3~$\upmu $m.
 It is observed that the charge collection time is a strong function of applied voltage.
 At 200~V, well above the depletion voltage of 155~V, it extends to about 0.5~$\upmu $s, which is longer than the 220~ns time interval between XFEL pulses.
 The right side of figure~\ref{fig:PlasmaTime} shows for the same sensor as function of the focussing of the laser light the current transients at a voltage of 500~V for the situation of $1.5\times 10^{5}$ photons of 12~keV.
 We conclude that the plasma effect is relevant for the operation of silicon sensors at the XFEL, and that operating voltages of about 500~V are required for 450~$\upmu $m thick sensors to assure a complete charge collection in-between XFEL pulses.

\begin{figure}
   \centering
	\includegraphics[width=7cm]{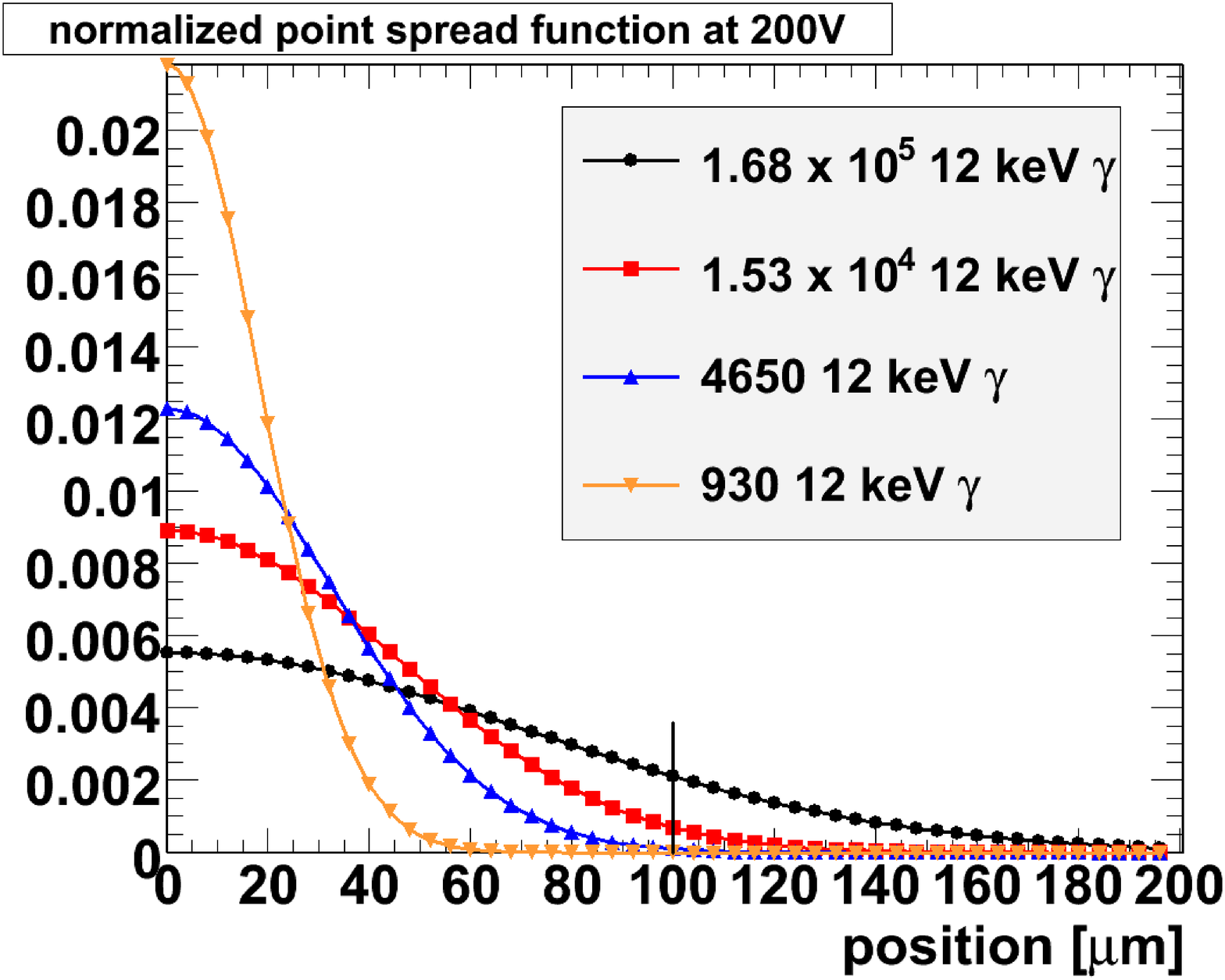}
	\includegraphics[width=7cm]{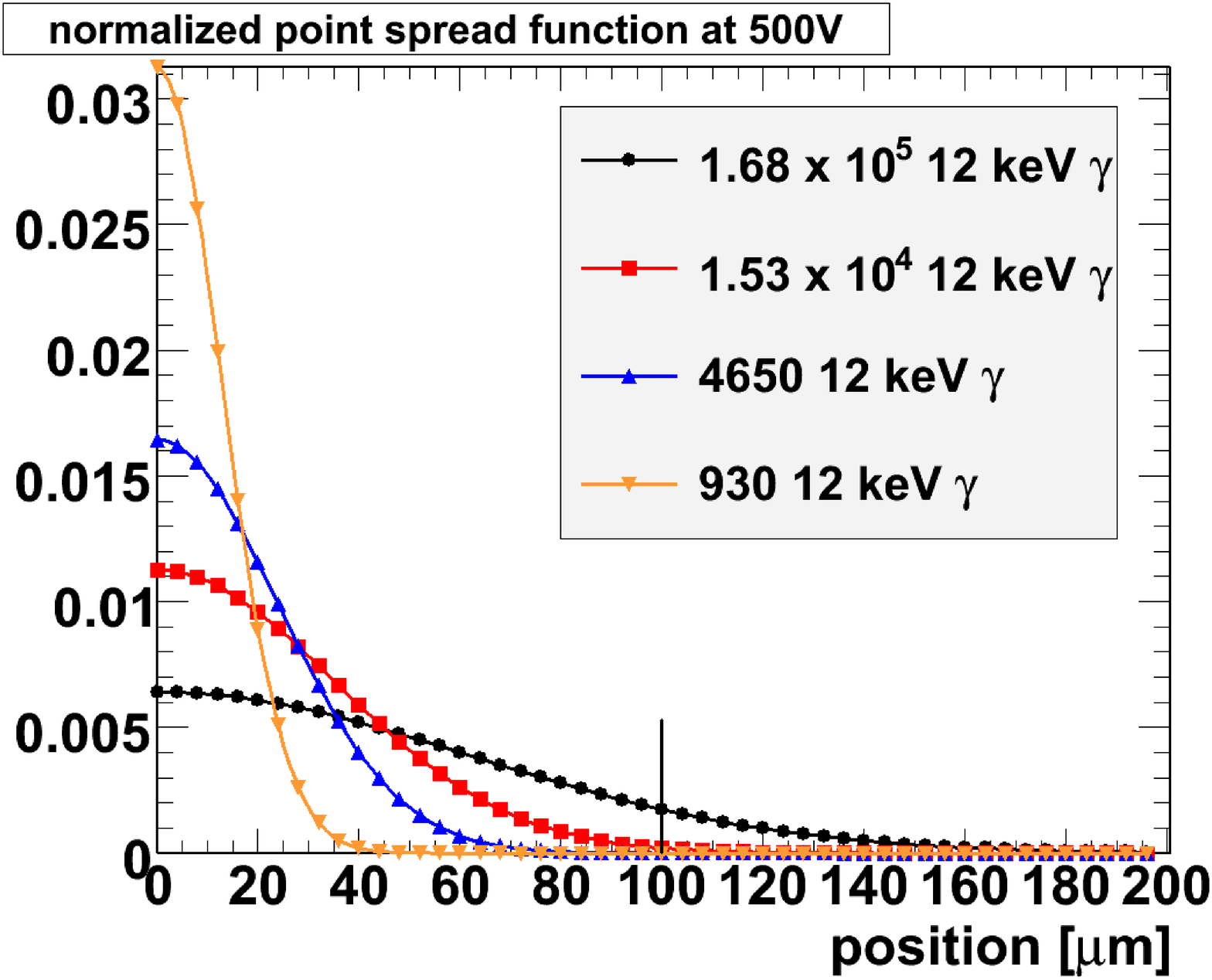}
   \caption{Normalized point spread functions for a 450~$\upmu$m silicon sensor with a depletion voltage of 155~V for different intensities of the 1015~nm laser focussed to a $\sigma $ of 3~$\upmu$m. Left: Applied voltage 200~V. Right: Applied voltage 500~V.}
  \label{fig:PlasmaSpace}
 \end{figure}

 By scanning the laser spot over the sensor and measuring the integrated charge on the individual read-out strips the point-spread function as function of the X-ray intensity and operating voltage has been determined.
 The results are shown in figure~\ref{fig:PlasmaSpace}.
 A strong dependence on intensity is observed.
 For a voltage of 200~V the rms width of the point-spread function increases from about 20~$\upmu $m for 930 to about 120~$\upmu $m for $1.7 \times  10^{5}$ X-ray photons.
 The corresponding values for 500~V are 20~$\upmu $m and 80~$\upmu $m.
 As for a number of experiments at the XFEL the precise measurement of the shape of high-intensity diffraction peaks is important, we conclude that an operation voltage well above 500~V should be possible for AGIPD.
 Thus in the sensor design an effort was made to reach a breakdown voltage approaching 1000~V.

\section{Radiation damage parameters}

 At the European XFEL radiation doses up to 1~GGy (SiO$_2$) from X-rays, highly non-uniformly distributed over the sensor, are expected.
 For such high doses no experience exists for high-ohmic structures like thick silicon sensors.
 Therefore we have undertaken the following program:
  \begin{itemize}
   \item Irradiate test structures from different vendors and with different crystal orientations to extract microscopic and macroscopic parameters due to X-ray-radiation damage.
   \item Irradiate sensors, measure their performance and use detailed TCAD simulations to understand the observations.
   \item Optimize the sensor design with TCAD simulations using the parameters obtained from the irradiated and non-irradiated test structures.
   \item Discuss the conclusions and the design with vendors, order  sensors and verify their performance.
  \end{itemize}

 The effects of X-ray-radiation damage are discussed in detail in \cite{Oldham:1999}.
 Here we only give a very short summary:
 In SiO$_2$ X-rays produce on average one $eh$ pair every 18~eV of deposited energy.
 A fraction of the $eh$ pairs, which strongly depends on ionization density and  electric field, will recombine.
 The remaining free charge carriers will move in the SiO$_2$ by diffusion and, if an electric field is present,  by drift.
 The mobility of electrons of about 20~cm$^2$/(V$\cdot $s) is much higher than the one for holes, which is less than $10^{-5}$~cm$^2$/(V$\cdot $s).
 Therefore, most of the electrons leave the SiO$_2$, whereas holes get trapped in the SiO$_2$, mainly in a layer of a few~nm depth close to the Si-SiO$_2$~interface, or form interface traps at the Si-SiO$_2$~interface.
% We note, that for a 500~nm thick SiO$_2$ a dose of 1~MGy produces $4 \times 10^{16}$~$eh$~pairs/cm${-2}$, compared to about $10^{15}$~cm${-2}$ surface states.
% Thus effects of saturation are certainly expected at high doses.
 We denote the density of oxide charges by $N_{ox}$, and the density of interface traps as function of their energy $E$ relative to the conduction band by $D_{it}(E)$ with units 1/(eV$\cdot $cm$^{2}$).

 The build-up of positive oxide charges and interface traps causes following effects in segmented $p^+n$~sensors:
 An accumulation layer forms at the Si-SiO$_2$~interface which can cause high-field regions close to the $p^+n$ junction resulting in a reduced breakdown voltage, an increase in the depletion voltage, charge losses close to the interface and an increased inter-pixel capacitance.
 The interface traps close to the middle of the silicon-band gap act as generation centers for surface currents, if they are exposed to an electric field.

 Test structures from 4 different vendors,
 Canberra \cite{Canberra},
 CiS \cite{CIS},
 Hamamatsu \cite{Hamamatsu}, and
 Sintef \cite{Sintef},
 were used for determining the parameters of X-ray-radiation damage.
 The test structures were fabricated on $n$-type silicon with crystal orientations $\langle 111 \rangle$ and $\langle 100 \rangle$, SiO$_2$ thicknesses between 250 and 750~nm, and with and without additional Si$_3$N$_4$ layers on top of the SiO$_2$.
 The MOS capacitors (MOS-C) for Canberra, CiS, and Hamamatsu were circular, with a diameter of 1.5~mm.
 The Sintef MOS-C was a rectangle of 1~mm $\times $ 3.5~mm.
 The Gate-Controlled Diodes (GCD) of CiS and Hamamatsu were circular with 1~mm diameter and 5 Al-gate rings of 50~$\upmu $m width on top of the insulator, each separated by 5~$\upmu $m.
 The GCD of Sintef was circular with a cental diode of 400~$\upmu $m diameter surrounded by a 210~$\upmu $m wide gate.
 The Canberra GCD was a finger structure with 100~$\upmu $m wide gates and 100~$\upmu $m wide diodes.
% The test structures used for determining the parameters of X-ray-radiation damage were circular MOS capacitors (MOS-C) with 1.5~mm diameter, and circular Gate-Controlled Diodes (GCD) with a central diode of 1~mm diameter and 5 Al-gate rings of 50~$\upmu $m width on top of the insulator each separated by 5~$\upmu $m.
% The test structures were fabricated by 4 different vendors on $n$-type silicon with crystal orientations $\langle 111 \rangle$ and $\langle 100 \rangle$, SiO$_2$ thicknesses between 300 and 700~nm, and with and without additional Si$_3$N$_4$ layers on top of the SiO$_2$.

 The irradiations have been performed in the "white" X-ray beam F4 at DORIS~III~\cite{Perrey:Thesis} with a mean energy of $\sim 12$~keV and dose rates between 1 and 200~kGy/s.
  \begin{figure}
   \centering
	\includegraphics[width=7.4cm]{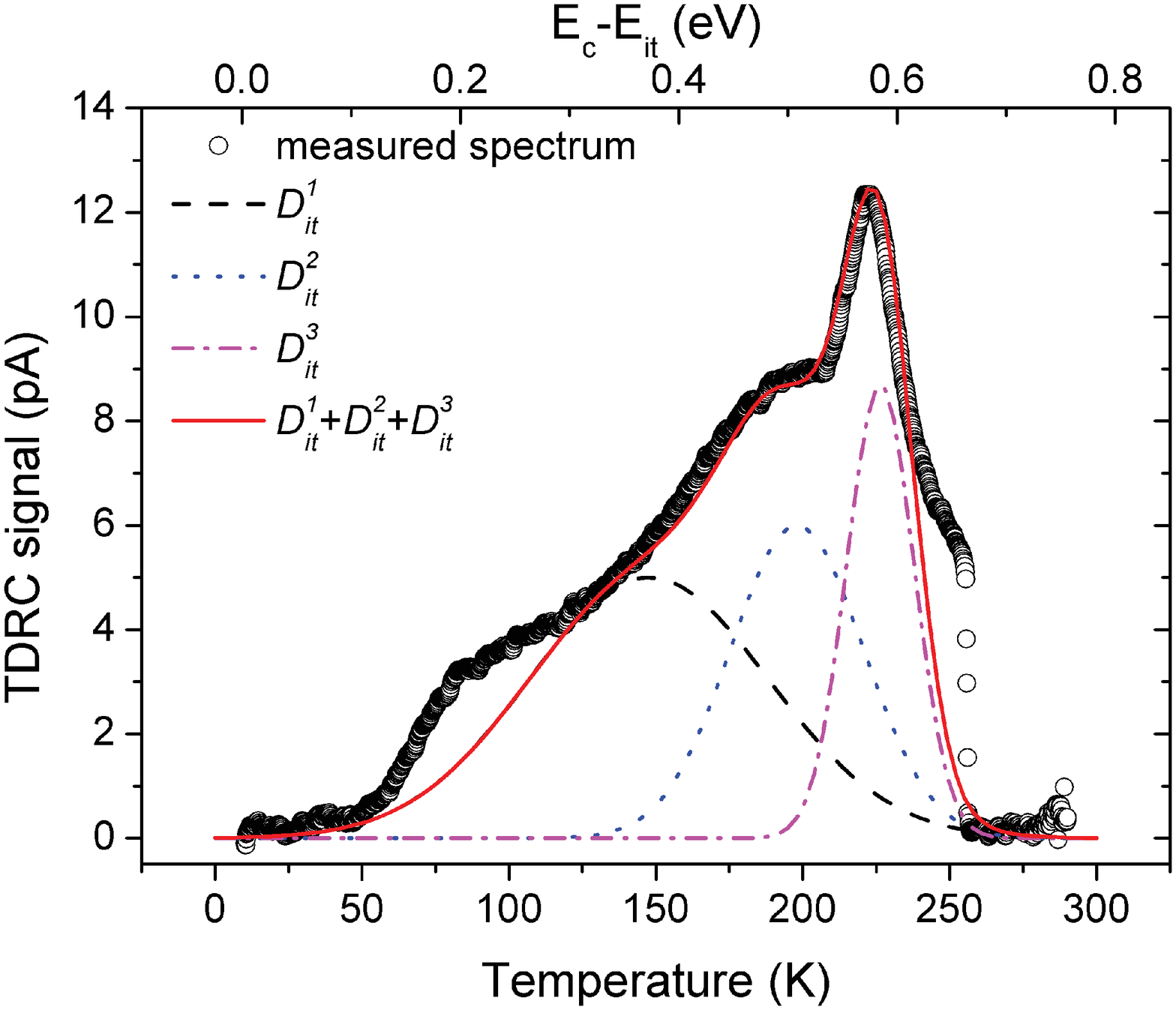}
	\includegraphics[width=7.4cm]{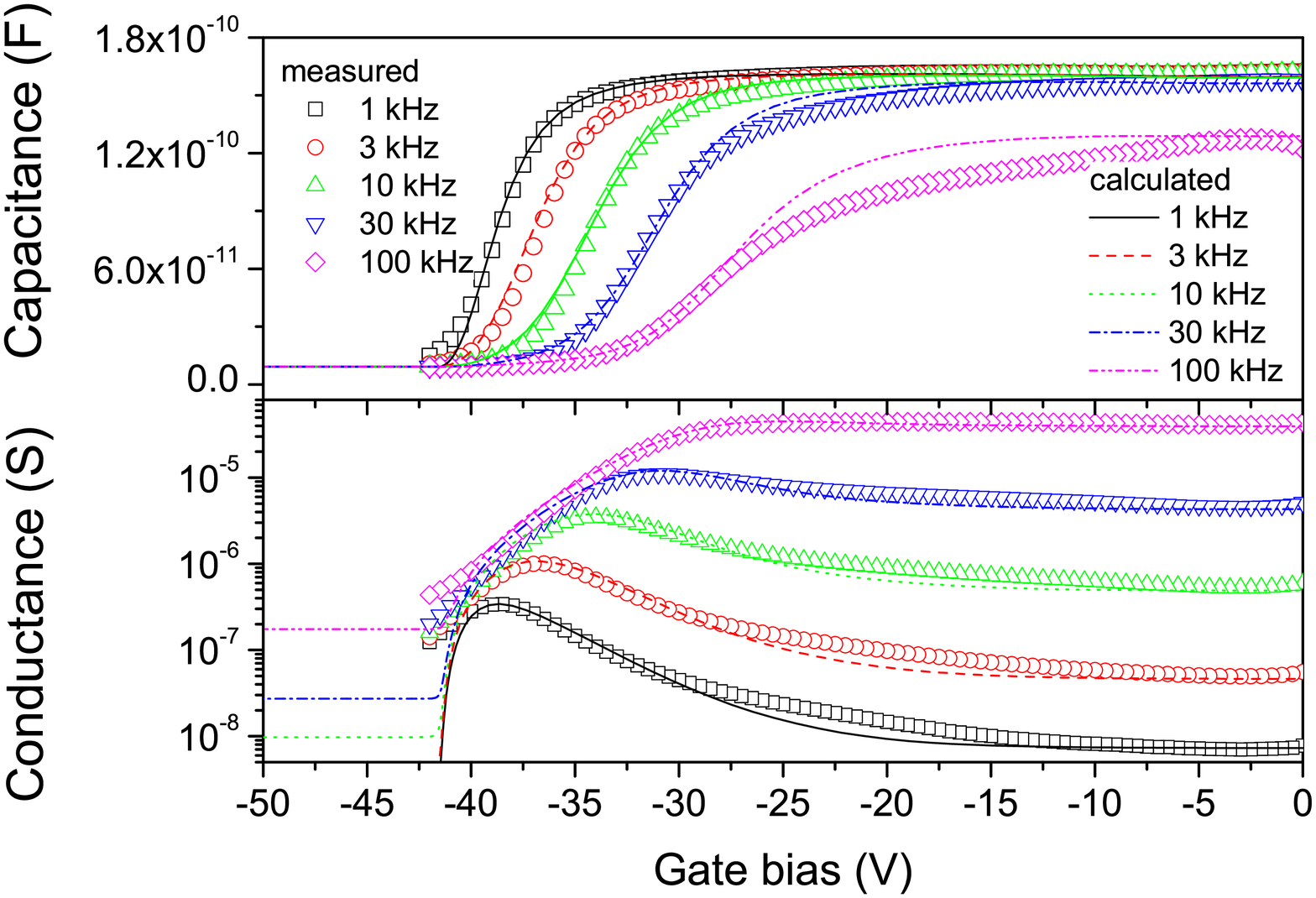}	
   \caption{Results from the measurement of a MOS capacitor irradiated to 5~MGy and annealed for 10~minutes at $80^\circ $C. Left: TDRC current with fit by 3 Gaussians.
   The top scale shows the approximate energy in the silicon-band gap measured from the conduction band.
   Right: C-V and G-V measurements for frequencies between 1 and 100~kHz.
   The curves show the results of the model used to determine $N_{ox}$.}
  \label{fig:MOS-C}
 \end{figure}

 For determining the interface-trap density $D_{it}(E)$ we have used the Thermal Dielectric Relaxation Current technique, TDRC \cite{Zhang:2011}:
 The MOS-C was brought in the state of electron accumulation (0~V bias), and cooled down to a temperature 10~K to freeze the electrons in the interface traps.
 Then the MOS-C was biased to deep depletion, heated up with a constant heating rate $\beta = 0.183$~K/s to 290~K, and the current $I_{TDRC}(T)$ due to the release of the trapped electrons was measured.
 As the temperature $T$ is directly related to the distance of the Fermi level from the conduction band, the interface-trap densities $D_{it}(E)$ can be directly extracted from $I_{TDRC}(T)$.
 Figure~\ref{fig:MOS-C} left shows the results for a MOS-C produced by CiS~\cite{CIS} (insulator thickness 350~nm SiO$_2$ plus 50~nm Si$_3$N$_4$, crystal orientation $\langle 100 \rangle$) irradiated to 5~MGy after annealing for 10~minutes at 80$^\circ $C.
 As shown in the figure, we used 3 Gaussians to parameterize $D_{it}(E)$.
 We note, that the method as used is only sensitive to electron traps with energies between the conduction band and $\sim $~0.6~eV below the conduction band, as the currents from the depletion region dominates for temperatures above $\sim $~250~K.
 An annealing for 10~minutes at 80$^\circ $C was required to obtain consistent results.
 As discussed below, the reason is the short annealing-time constant already at room temperature of some of the interface traps.
 In addition, the description with 3 states is certainly not unique.

\begin{figure}
   \centering
   \includegraphics[width=7.2cm]{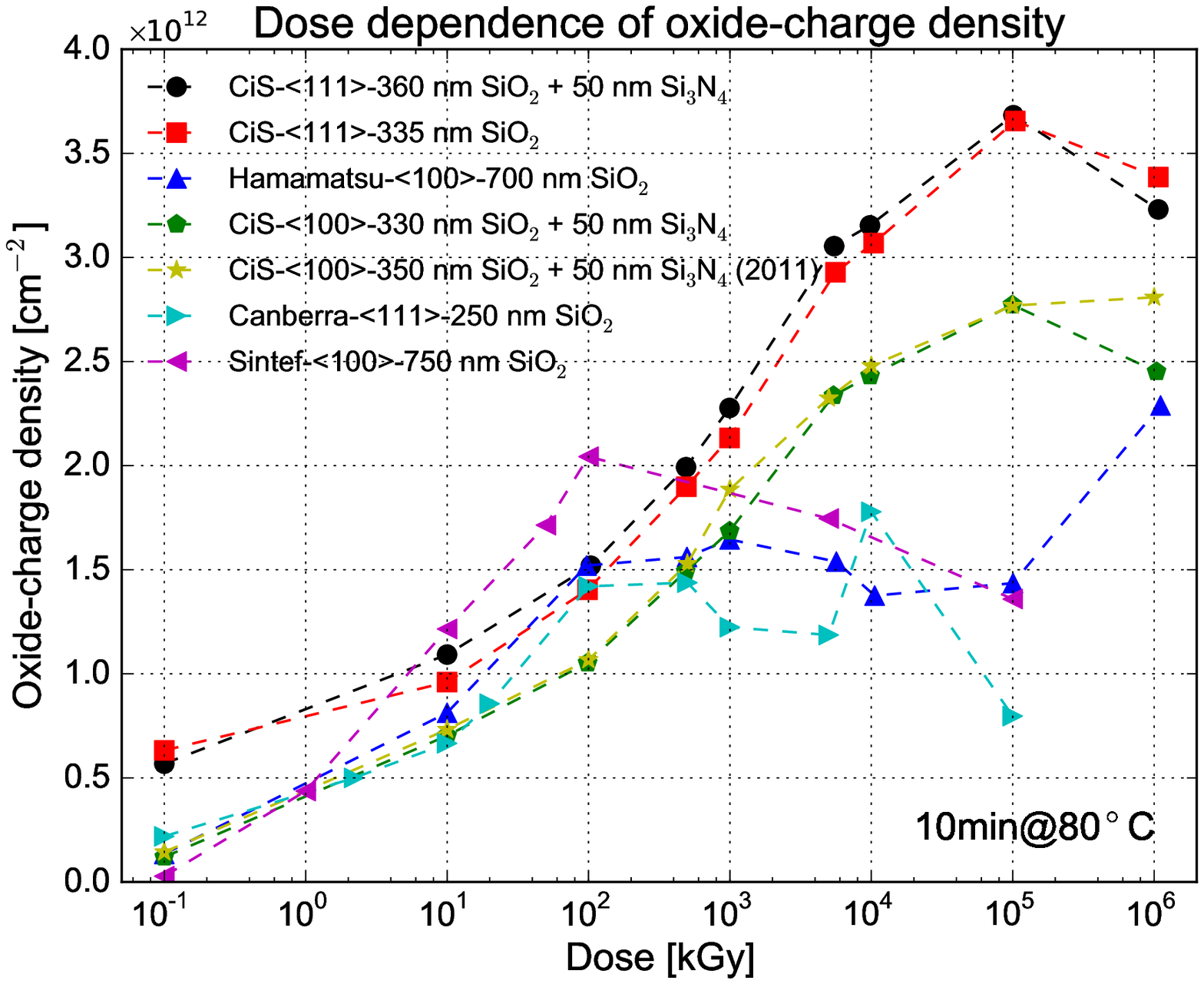}
	\includegraphics[width=6.8cm]{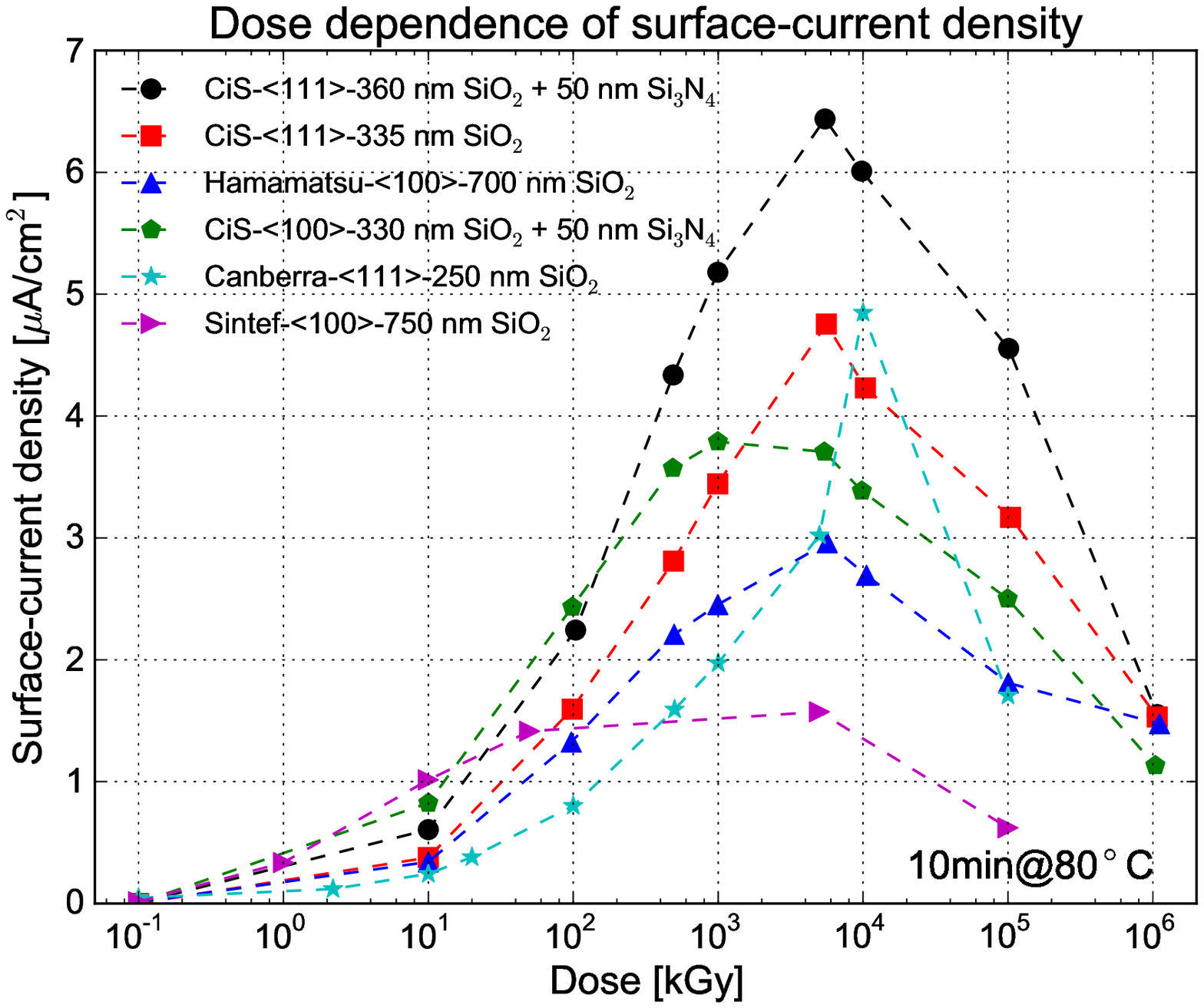}
   \caption{Dose dependence of X-ray radiation damage obtained from the test structures from different vendors.
   Left: Oxide-charge density $N_{ox}$.
   Right: Surface-current density $J_{surf}$ scaled to $20^\circ$C.}
  \label{fig:DoseDependence}
 \end{figure}

 In order to determine the oxide-charge density $N_{ox}$, but also to verify that the description of the interface-trap density $D_{it}(E)$ is valid, C/G-V measurements on the MOS-C were made.
 Figure~\ref{fig:MOS-C} right shows an example of such a measurement for frequencies between 1 and 100~kHz.
 The oxide capacitance $C_{ox}$ calculated from the insulator thickness and the MOS-C area is 160~pF, the capacitance expected in inversion $C_{inv} \sim 10$~pF, and the flat-band capacitance $C_{flat} \sim 31$~pF.
 The measurements show a strong frequency dependence of both C-V and G-V curves due to the presence of interface traps.
 The frequency below which the interface traps contribute to the capacitance depends on their energy in the Si-band gap, their cross sections for charge carriers and the band bending at the Si-SiO$_2$ interface.
 Following~\cite{Nicollian:1982} an R-C model has been developed to describe the frequency dependence of the C-V and G-V curves~\cite{Zhang:2011}.
 The oxide-charge density $N_{ox}$ only shifts the curves along the V~axis, and $N_{ox}$ has been determined by shifting the C/G~curves obtained from the model using the measured TDRC spectra, until the data are approximately described.
 The results, shown as solid lines in figure~\ref{fig:MOS-C} right, demonstrate that a fair description of the measurements has been achieved.
 The analysis described above has been performed for the different MOS capacitors irradiated to different X-ray doses, and the results are shown in figure~\ref{fig:DoseDependence} left.
 It is observed that, in spite of the different technologies, oxide thicknesses and crystal orientations, the trends of $N_{ox}$ as function of X-ray dose are similar: $N_{ox}$ increases up to dose values of 10 to 100~MGy and then saturates, with  saturation values between 2 and $4 \times 10^{12}$~cm$^{-2}$.

 For determining the surface-current density, I-V measurements on Gate-Controlled Diodes (GCD)~\cite{Grove:1966, Grove:1967, Fitzgerald:1968} were made.
 The diodes were biased to --12~V and the diode current as function of voltage on the first gate measured.
 The surface current $I_{surf}$ is obtained from the difference of the current when the gate is in depletion and when it is in accumulation~\cite{Zhang:2011}, and the surface current density $J_{surf}$ by dividing $I_{surf}$ by the gate area, thus ignoring gate-length effects \cite{Pierret:1974, DallaBetta:2000}. Such effects are presently under study. They indicate that for the longer gates (Canberra and Sintef) the surface-current densities may be underestimated by 25 to 50~\%.
 The results for $J_{surf}$  are shown in figure~\ref{fig:DoseDependence} right.
 As for $N_{ox}$, the general trend for all samples is similar:
 An increase up to X-ray doses of 1 to 10~MGy, then however a decrease up to 1~GGy, the highest dose measured.
 The maxima of $J_{surf}$ are between 3 and 6~$\upmu$A/cm$^2$.
 The reason for the decrease it not yet understood, but it may well be related to gate-length dependent effects.
% It has, however to a lesser extend, also been observed in the dose dependence of the interface traps and the dark current of strip sensors.
 We have also observed a similar dose dependence of the dark current for irradiated micro-strip sensors.

   \begin{figure}
   \centering
    \includegraphics[width=7cm]{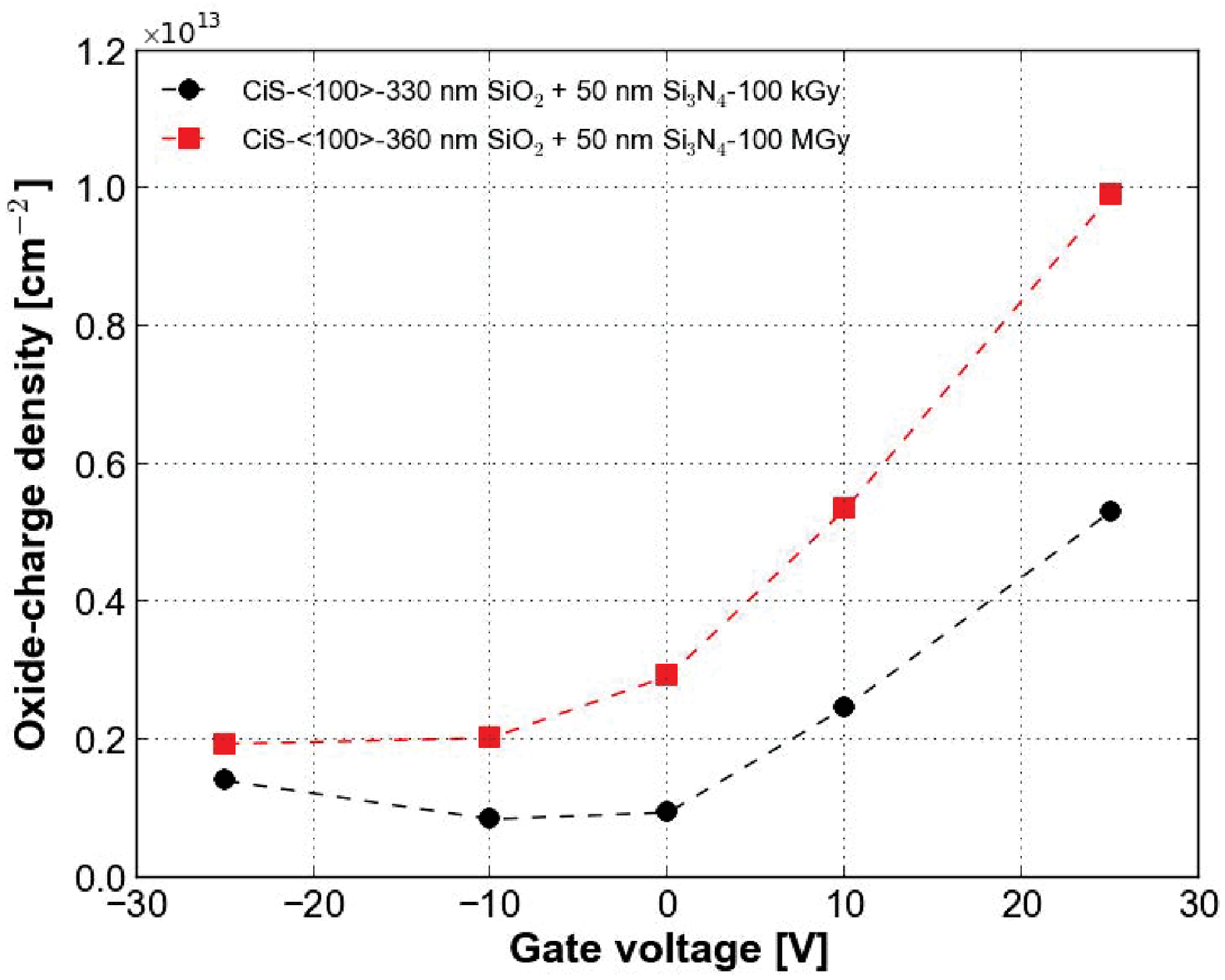}
	\includegraphics[width=7cm]{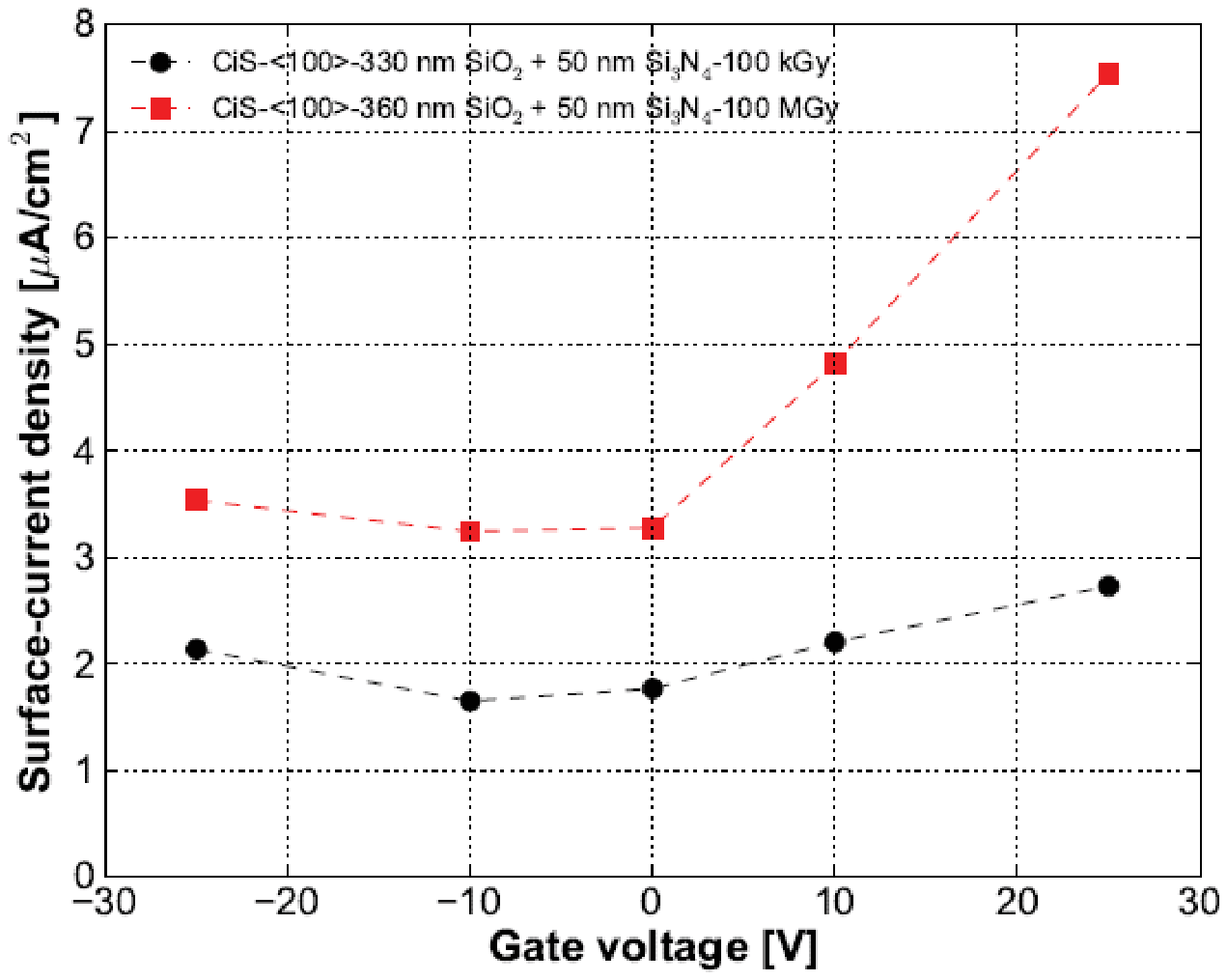}
   \caption{Dependence of X-ray-radiation damage on the voltage applied during irradiation.
   Left: Oxide-charge density $N_{ox}$.
   Right: Surface-current density $J_{surf}$.}
  \label{fig:EDependence}
 \end{figure}
 The results presented so far, have been obtained from test structures which were not biased during irradiation.
 In order to investigate the impact of the applied bias voltage, we have also irradiated MOS capacitors and gate-controlled diodes from CiS~\cite{CIS} (350~nm SiO$_2$, 50~nm Si$_3$N$_4$, crystal orientation $\langle 100 \rangle$) to 100~kGy and 100~MGy with voltages between --~25 and +25~V applied during irradiation.
 The results for $N_{ox}$ and $J_{surf}$ are shown in figure~\ref{fig:EDependence}.
 For negative voltages, where the electric field points from the silicon to the Al gate, $N_{ox}$ and $J_{surf}$ are approximately independent of voltage.
 For positive voltages however, there is a significant increase.
 This observation agrees with expectations, as the positive field drives the holes towards the Si-SiO$_2$ interface.
 We note that in a $p^+n$ sensor the situation corresponds to a negative voltage, and thus no significant increase with respect to the zero-field situation is expected.

   \begin{figure}
   \centering
	\includegraphics[width=7cm]{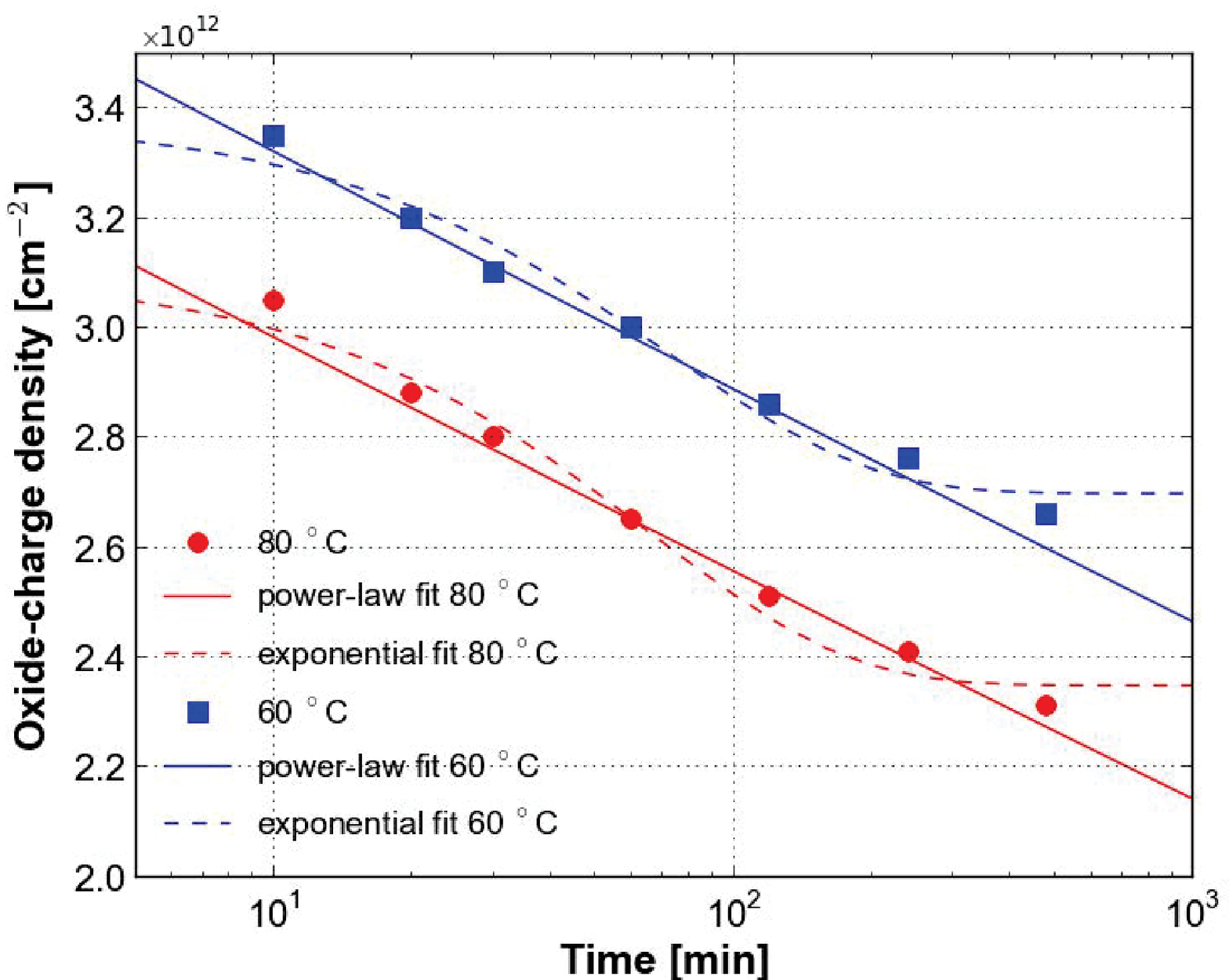}
	\includegraphics[width=7cm]{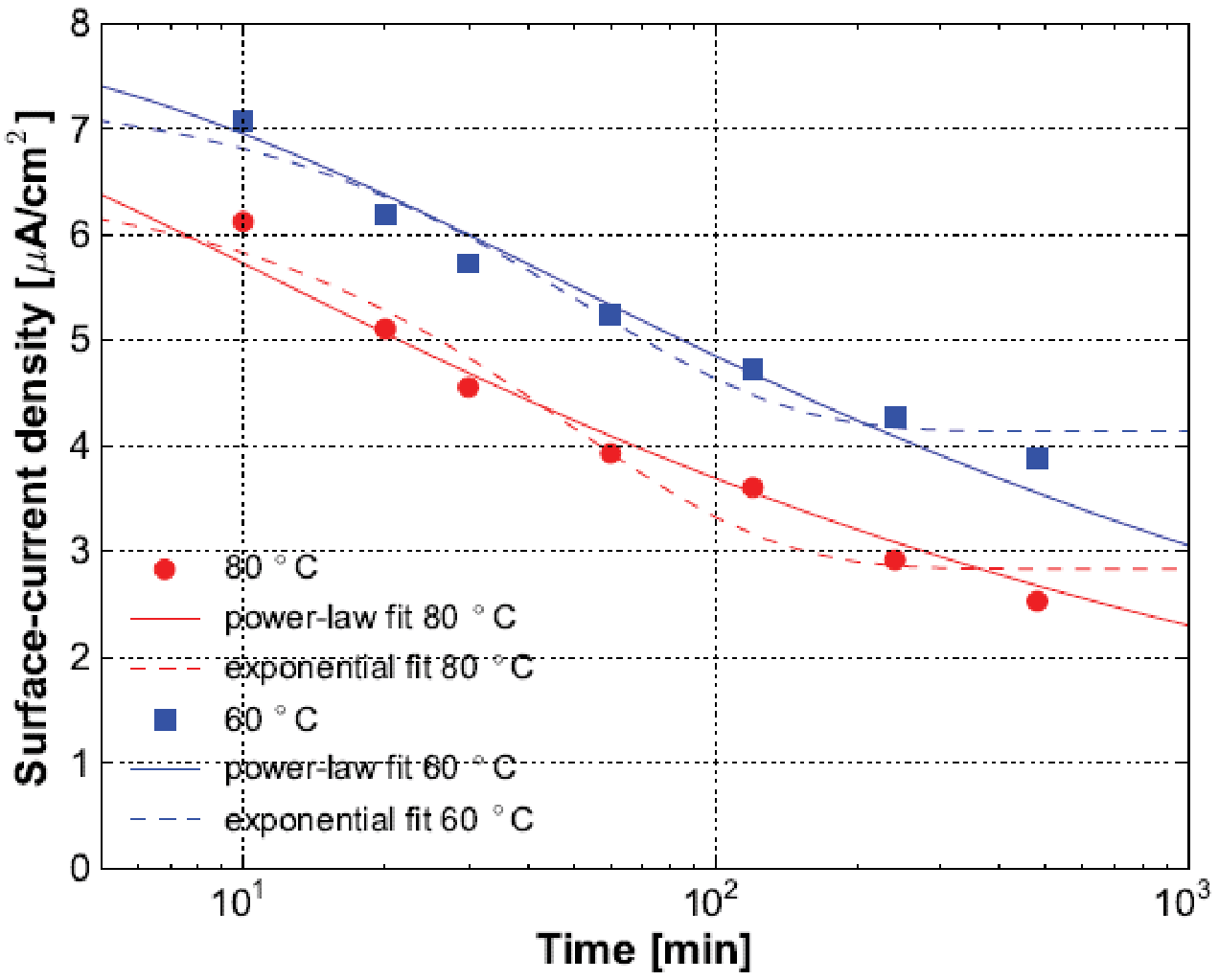}
   \caption{Annealing of the X-ray-radiation damage at 60 and 80$^\circ $C. The curves are fits by an exponential (dashed) and a modified power law (continuous).
   Left: Oxide-charge density $N_{ox}$.
   Right: Surface-current density $J_{surf}$.}
  \label{fig:ADependence}
 \end{figure}

 Finally, the annealing of $N_{ox}$ and of $J_{surf}$ has been studied for the CiS structures described above at 60 and 80$^\circ $C.
 The results for test structures irradiated to 5~MGy are shown in figure~\ref{fig:ADependence}.
 A fit of an exponential function to the measured time dependence, which assumes a constant annealing probability, does not describe the data.
 However, good descriptions were obtained by
  $N_0 \cdot (1 + t/t_0)^n$ with
  $t_0(T) = t_0^* \cdot exp(\Delta E/(k_B \cdot T))$
 with the free parameters $N_0$, $n$, $t_0^*$, and activation energy $\Delta E$.
 $t_0(T)$ is the characteristic time constant at the temperature $T$, and $k_B$ the Boltzmann constant.
 This time dependence is expected by the "tunnel-anneal model"~\cite{Oldham:1988} for $N_{ox}$, and by the "two-reaction model"~\cite{Reed:1987} for $J_{surf}$.
 For the parameters found we refer to~\cite{Zhang:2012}.
 Using these models we  estimate a time for 50~\% annealing at $20 ^\circ $C of 3 years for $N_{ox}$, and of 5 days for $J_{surf}$.
 We however should warn that the extrapolation from the  measurements at 60 and 80$^\circ $C to  $20 ^\circ $C results in large uncertainties.
 Nevertheless, the time constant for $J_{surf}$ is short and explains why annealing was necessary for obtaining consistent results.

 To summarize:
 X-ray irradiation results in an increase of the oxide-charge density and of the density of interface traps, which are responsible for the generation of surface current.
 For doses in the range between 10 and 100~MGy both oxide-charge density and surface-generation current saturate.
 The increase of both oxide-charge density and surface-generation current depends on the value and the direction of the electric field during the irradiation.
 However for $p^+n$~sensors the electric field points in the favorable direction resulting in values similar to the situation without applied electric field.
 For both oxide-charge density and surface-generation current annealing has been observed.
 Its time dependence cannot be described by an exponential, but by a modified power law.

 \section{Impact of radiation damage on segmented $p^+n$ sensors}

 The influence of X-ray-radiation damage on segmented $p^+n$~sensors has been investigated both experimentally and by TCAD simulations \cite{Zhang1:2012, Schwandt1:2012}.
 As a result of the positive charges in the oxide and at the Si-SiO$_2$~interface, an electron-accumulation layer forms at the interface, or, if already present for the non-irradiated sensor, its width increases \cite{Poehlsen:2012}.
 This results in an increase of the depletion voltage for segmented sensors of typically 10~V, which however is of little relevance for the standard operation.

 More important is, that the interface traps cause an increase of the dark current by several orders of magnitude.
 The current is given by the product of the surface-current density $J_{surf}$ and the area of the depleted Si-SiO$_2$~interface, which depends on the difference of the distance between the $p^+$~implants minus the width of the electron-accumulation layer.
 As the accumulation layer shrinks with increasing bias voltage, the dark current does not saturate when the sensor is depleted, but continues to rise approximately linear with voltage.
 In order to limit the maximum dark current, the gap between the $p^+$~implants should be small, which however results in an increase of the inter-pixel capacitance.
 For the design of the AGIPD sensor a gap of 20~$\upmu $m has been chosen.
  \begin{figure}
   \centering
	\includegraphics[width=4.8cm]{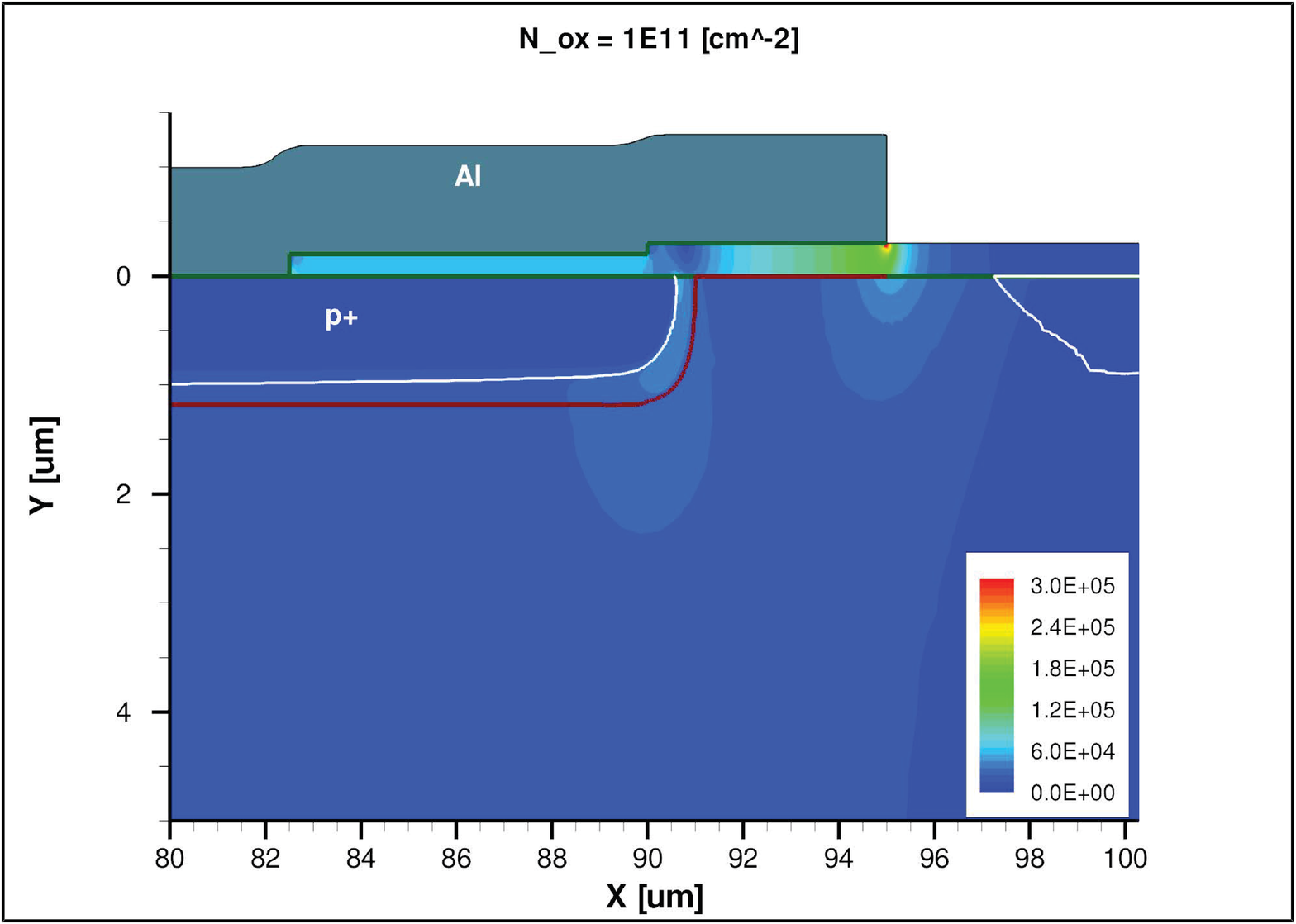}
	\includegraphics[width=4.8cm]{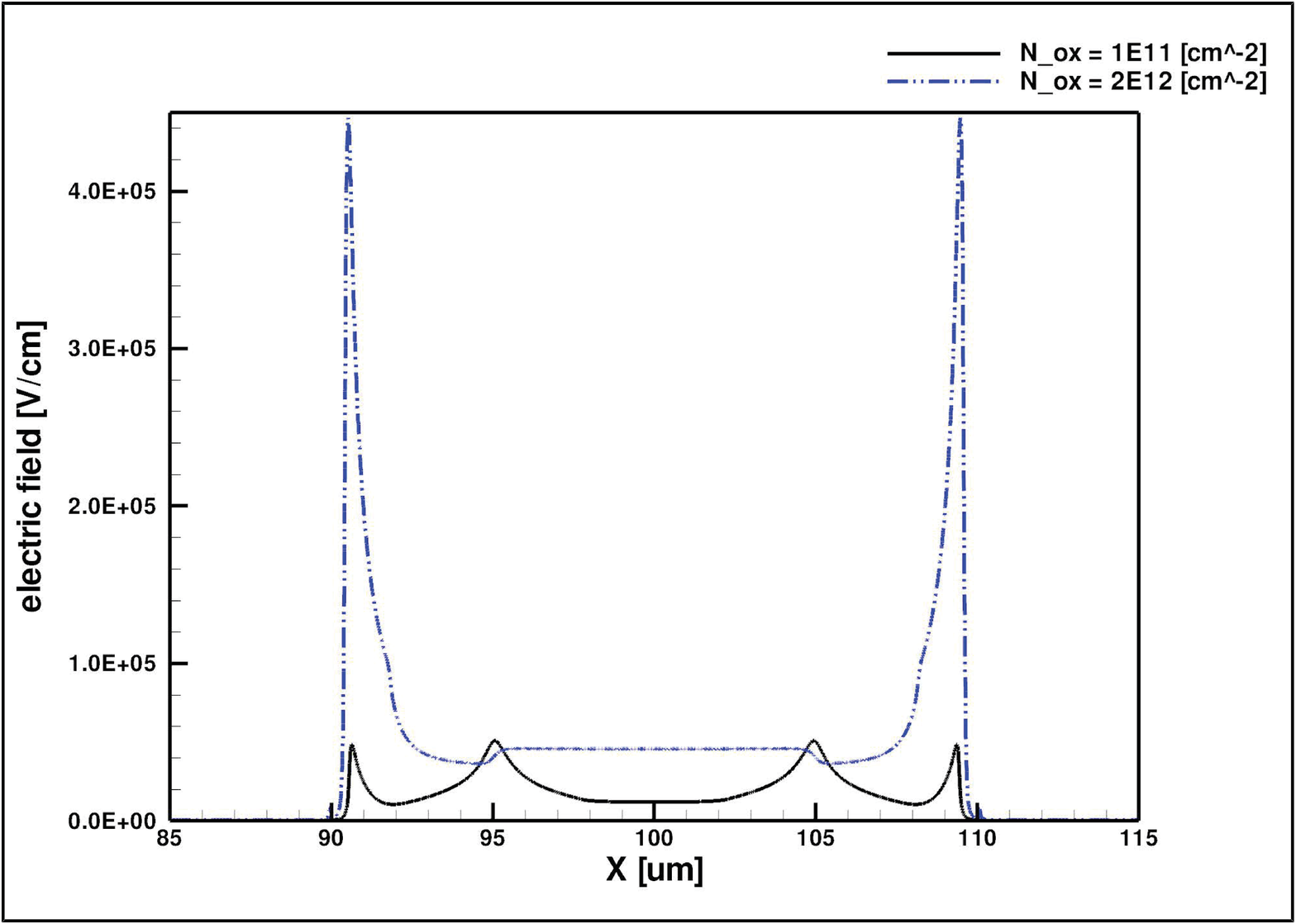}	
	\includegraphics[width=4.8cm]{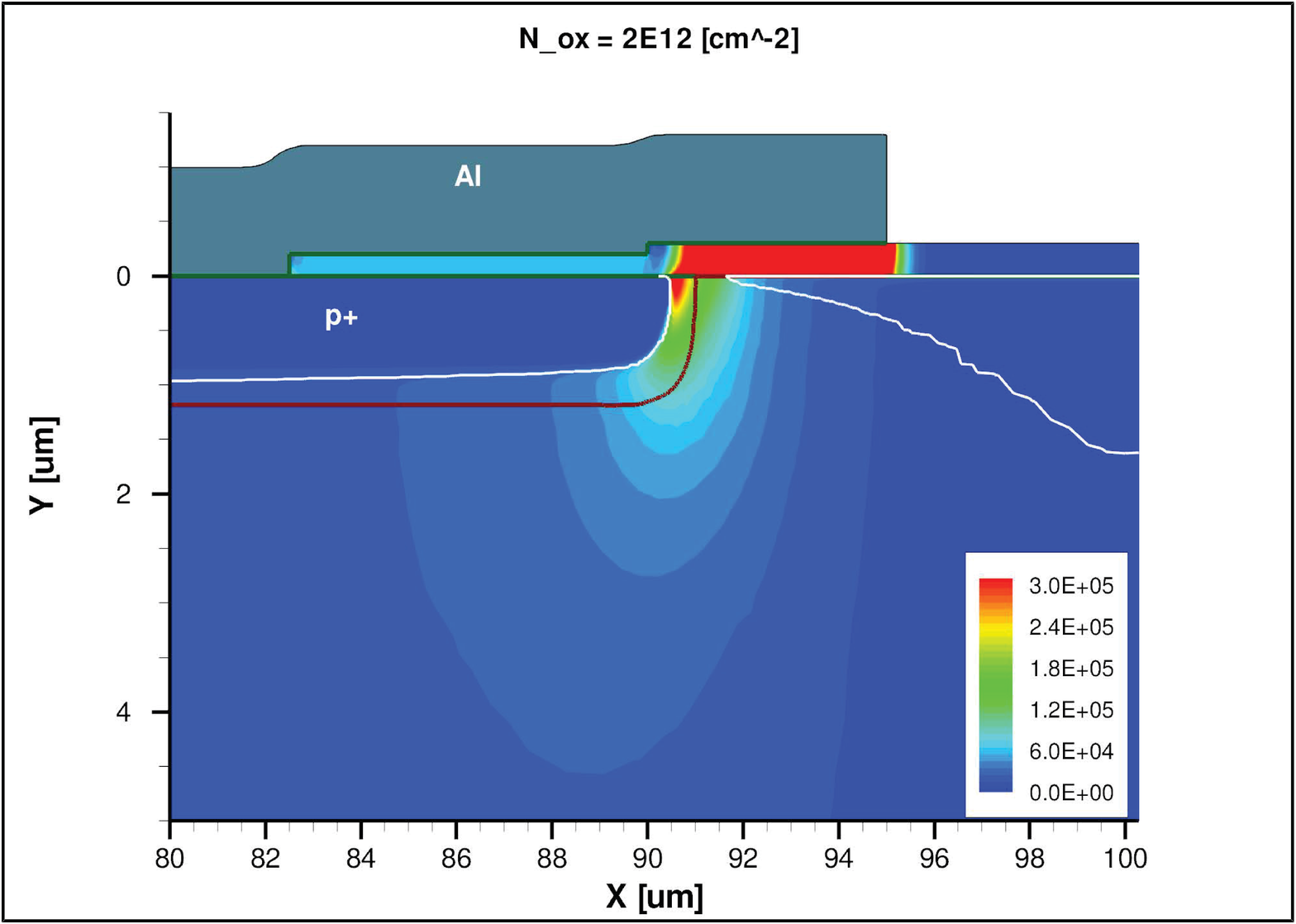}
     \caption{Influence of the oxide-charge density $N_{ox}$ on the electric field close to the Si-SiO$_2$ interface.
     Left: 2D field distribution for $N_{ox} = 10^{11}$~cm$^{-2}$.
     Right: 2D field distribution for $N_{ox} = 2 \times 10^{12}$~cm$^{-2}$.
     Center: Electric field in the silicon 10~nm from the  Si-SiO$_2$ interface for the two values of $N_{ox}$.}
  \label{fig:Efield}
 \end{figure}

 The most important effect of the electron-accumulation layer is, that a high-field region appears at its edge below the aluminium overhang and/or at the corner of the $p^+$ implant.
 This is demonstrated in figure~\ref{fig:Efield}, where a simulation of the electric field 10~nm below the Si-SiO$_2$ interface is shown for values of $N_{ox}$ of $10^{11}$ and $2 \times 10^{12}$~cm$^{-2}$.
 The corresponding values of the maximal fields are 50 and 450~kV/cm.
 Thus X-ray damage results in a significant reduction of the breakdown voltage, and, as discussed in \cite{Schwandt1:2012} the optimization of the guard-ring structure is very different for irradiated and non-irradiated sensors.
 According to the simulations 15 guard rings are required to reach breakdown voltages approaching 1000~V for X-ray doses in the entire range between 0 and 1~GGy, the thickness of the SiO$_2$ has to be  about 250~nm, and a deep $p^+$ implant of about 2~$\upmu $m is advantageous.

 The electron-accumulation layer also influences the lateral extension of the depletion layer at the edge of the sensor: the higher the electron density the smaller the lateral extension.
 In the design of the AGIPD sensor this has been taken into account by an $n^+$~implantation between the outer guard ring and the scribe line.

\section{Charge losses close to the Si-SiO$_2$ interface}

 Given the large number of up to $10^5$ X-ray photons per XFEL bunch in a single pixel and the short time interval of 220~ns between XFEL pulses, it is important to verify that no charges are stored in low-field regions, which can cause pile-up.
 This has been studied by the TCT set-up described already in section~2 for the study of the plasma effect.
 Focussed light with 3.5~$\upmu$m penetration depth in silicon was used to produce $eh$ pairs close to the Si-SiO$_2$~interface and the time resolved pulses were recorded.
 This study has revealed that actually significant charge losses close to the interface occur, however the overall losses for X-ray photons entering the $n^+$ side of the sensors are at the per mille level and can be ignored.
 It  could be shown that the charges are lost in the accumulation layer, where they disperse on time scales below 1~ns over a large part of the sensor, and thus do not cause any danger for pile-up.
 During these studies it has been found that the measurement of charge losses is a tool to investigate the properties of both hole- and electron-accumulation layers at the Si-SiO$_2$~interface and their dependencies on the electric boundary conditions on the sensor surface, which in turn are influenced by the biasing history and the humidity \cite{Poehlsen:2012, Poehlsen1:2012}.

\section{Summary}

 The challenges for $p^+n$ silicon-pixels sensors posed by the high intensity and high bunch rate of the European XFEL have been studied experimentally.

 The high instantaneous density of X-rays causes the so called plasma effect, which results in longer signals and an increased spread of the collected charge.
 The effect decreases with increased operating voltage.
 For a 450~$\upmu $m thick sensors a voltage above 500~V is recommended.
 For some applications operating voltages approaching 1000~V appear desirable.

 The main challenge is X-ray-radiation damage, which causes an increase in oxide charges and interface traps.
 It is found that both saturate at densities of a few~10$^{12}$~cm$^{-2}$.
 The additional interface traps cause an increase by several orders of magnitude of the surface-generation current at the Si-SiO$_2$ interface, and positive oxide charges lead to an electron-accumulation at the Si-SiO$_2$~interface.
 In order to reduce the surface-generation current, the gap between the $p^+$~implants should be made small, compatible with the increased readout noise due to the increased inter-pixel capacitance.
 The main effect of the electron-accumulation layer is the generation of high-field regions at the edge of the $p^+$ implant and below the aluminium overhang.
 This makes the design of a guard-ring structure for voltages approaching 1000~V a major challenge, and the optimized technological and layout parameters are quite different compared to non-irradiated sensors.

 Based on the measurements described in this article and on extensive TCAD simulations the pixel sensor for the AGIPD project has been designed and is presently fabricated in industry.

\section*{Acknowledgements}

  This work was performed within the AGIPD Project which is partially supported by the XFEL-Company.
  We would like to thank the AGIPD colleagues for the excellent collaboration.
  Support was also provided by the Helmholtz Alliance "Physics at the Terascale" and the German Ministry of Science, BMBF, through the Forschungsschwerpunkt "Particle Physics with the CMS-Experiment".
  J.~Zhang is supported by the Marie Curie Initial Training Network "MC-PAD", and I.~Pintilie gratefully acknowledges the financial support from the Romanian Authority for Scientific Research through the Project PCE~72/5.10.2011.
%  We also are thankful to H.~Spieler for essential advice and enlightening discussions.


\section*{References}

\begin{thebibliography}{9}

%[1] M. Altarelli et al., Technical Design Report, ISBN 978-3-935702-17-1 (2006).
%[2] Th. Tschentscher et al, TECHNICAL NOTE XFEL.EU TN-2011-001 (2011). DOI:10.3204/XFEL.EU/TR-2011-001.
%[3] H. Graafsma, 2009 JINST 4 P12011, DOI: 10.1088/1748-0221/4/12/P12011.
%[4] B. Henrich et al., Nucl. Instr. and Meth. A 663 2011 S11-14, DOI: 10.1016/j.nima.2010.06.107.
%[5] AGIPD: http://hasylab.desy.de/instrumentation/detectors/projects/agipd/index_eng.html.
%[6] J. Becker et al., Nucl. Instr. and Meth. A (2010), DOI:10.1016/j.nima.2010.01.082.
%[7] J. Zhang et al., J. Synchrotron Rad. (2012). 19, 340-346, DOI: 10.1107/S0909049512002348.
%[8] J. Zhang et al., 2011 JINST 6 C11013, DOI: 10.1088/1748-0221/6/11/C11013.
%[9] J. Zhang et al., Investigation of X-ray Induced Radiation Damage at the Si-SiO2 Interface of Silicon Pixel Sensors for the European XFEL, to be published in Proceedings of the 14th International Conference on Radiation Imaging Detectors 1 - 5 July 2012, Figueira da Foz, Portugal.
%[10] T. Pöhlsen, et al., Charge losses in segmented silicon sensors at the Si-SiO2 interface, arXiv:1207.6538v1 , submitted to Nucl. Instr. and Meth. A.
%[11] J. Schwandt et al., 2012 JINST 7 C01006, DOI: 10.1088/1748-0221/7/01/C01006.
%[12] J. Schwandt et al., Design of the AGIPD Sensor for the European XFEL, to be published in Proceedings of the 14th International Conference on Radiation Imaging Detectors 1 - 5 July 2012, Figueira da Foz, Portugal.

 \bibitem{XFEL}
  M.~Altarelli et al. (Eds.),
   \emph{XFEL: The European X-Ray Free-Electron Laser, Technical Design Report}, Preprint DESY 2006-097, DESY, Hamburg 2006, ISBN 978-3-935702-17-1, and \url{http://www.xfel.eu/de/}.

 \bibitem{Tschentscher:2011}
   Th.~Tschentscher et al., TECHNICAL NOTE XFEL.EU TN-2011-001 ~2011,
   DOI:~10.3204/XFEL.EU/TR-2011-001.

 \bibitem{Graafsma:2009}
   H. Graafsma, 2009~JINST~4~P12011~2011,
    DOI:~10.1088/1748-0221/4/12/P12011.

 \bibitem{Henrich:2011}
   B. Henrich et al., Nucl. Instr. and Meth.~A~663~(2011)~S11-14,
    DOI:~10.1016/j.nima.2010.06.107.

 \bibitem{Henrich1:2011}
   B.~Henrich et al., Nucl. Instr. and Meth. A~500~Suppl.~1(2011)~S11-S14,
    DOI:~10.1016/j.nima.2010.06.107.

 \bibitem{AGIPD}
   \url{http://hasylab.desy.de/instrumentation/detectors/projects/agipd/index_eng.html}.

 \bibitem{Tove:1967}
   P.A.~Tove and W.~Seibt,  Nucl. Instr. and Meth. 51~(1967)~261.

 \bibitem{Becker:2010}
  J. Becker et al., Nucl. Instr. and Meth. A~615~(2010)~230-236,
   DOI:~10.1016/j.nima.2010.01.082.

 \bibitem{Becker:Thesis}
  J.~Becker, \emph{Signal development in silicon sensors used for radiation detection}, PhD thesis, Universit\"at Hamburg, DESY-THESIS-2010-33~(2010).

 \bibitem{Zhang:2012}
  J. Zhang et al., J.~Synchrotron~Rad.~19~(2012)~340-346,
   DOI:~10.1107/S0909049512002348.

 \bibitem{Zhang:2011}
  J. Zhang et al., 2011~JINST~6~C11013,
   DOI:~10.1088/1748-0221/6/11/C11013.

 \bibitem{Zhang1:2012}
  J. Zhang et al., 2012~JINST~7~C12012,
  DOI:~10.1088/1748-0221/7/12/C12012.

%   \emph{Investigation of X-ray Induced Radiation Damage at the Si-SiO$_2$ Interface of Silicon Pixel Sensors for the European XFEL},
%    to be published in Proceedings of IWORID2012, the 14th International Workshop on Radiation Imaging Detectors, accepted by JINST, and arXiv:1210.0427.

 \bibitem{Poehlsen:2012}
  T. Poehlsen, et al.,
%  \emph{Charge losses in segmented silicon sensors at the Si-SiO$_2$ interface},
   Nucl. Instr. and Meth.~A~700~(2013) 22-39,
    DOI:~10.1016/j.nima.2012.10.063.

 \bibitem{Schwandt:2012}
  J. Schwandt et al., 2012~JINST~7~C01006,
   DOI:~10.1088/1748-0221/7/01/C01006.

 \bibitem{Schwandt1:2012}
  J. Schwandt et al.,
   \emph{Design of the AGIPD Sensor for the European XFEL},
    to be published in Proceedings of IWORID2012, the 14th International Workshop on Radiation Imaging Detectors, accepted by JINST, and arXiv:1210.0430.

 \bibitem{Gaertner:2010}
  J.~Becker et al., Nucl. Inst. Meth. A 624(3) (2010) 716-727, DOI:~10.1016/j.nima.2010.10.010.

 \bibitem{Oldham:1999}
   T.R~Oldham,
    \emph{Ionizing Radiation effects in MOS Oxides},
    \emph{World Scientific Publishing Co.} (1999).

 \bibitem{Canberra}
   Canberra Industries, Inc.,
   \url{http://www.canberra.com}.
%      \emph{http://www.canberra.com}.

 \bibitem{CIS}
  CiS Forschungsinstitut für Mikrosensorik und Photovoltaik GmbH,
%  Erfurt, Germany:
   \url{http://www.cismst.org}.

 \bibitem{Hamamatsu}
  Hamamatsu Photonics,
   \url{http://www.hamamatsu.com}.
%  \emph{http://www.hamamatsu.com}.

 \bibitem{Sintef}
  SINTEF ICT,
   \url{http://www.sintef.no}.
%  \emph{http://www.sintef.no/}.

 \bibitem{Perrey:Thesis}
   H.~Perrey,
    \emph{Jets at Low Q$^2$ at HERA and Radiation Damage Studies for Silicon Sensors for the XFEL}, PhD thesis, Universität Hamburg, DESY-THESIS-2011-021 (2011).

  \bibitem{Pierret:1974}
   R.F.~Pierret, Solid-State Electronics 17 (1974) 1257-1269.

  \bibitem{DallaBetta:2000}
   G.-F.~Dalla Betta et al., Proceedings of the 2000 International Conference on Microelectronic Test Structures, ICMTS 2000, DOI:~10.1109/ICMTS.2000.844410.

 \bibitem{Nicollian:1982}
   E.H.~Nicollian and J.R.~Brews,
    \emph{MOS (Metal Oxide Semiconductor) Physics and Technology}, New York, Wiley-Interscience, 1982.

 \bibitem{Grove:1966}
   A.S.~Grove and D.J.~Fitzgerald, Solid-State Electronics~9~(1966)~783.

 \bibitem{Grove:1967}
  A.S.~Grove,
   \emph{Physics and Technology of Semiconductor Devices}, John Wiley \& Sons (1967).

 \bibitem{Fitzgerald:1968}
   D.J.~Fitzgerald and A.S.~Grove, Surface Sience~9~(1968)~347.

 \bibitem{Oldham:1988}
   T.R.~Oldham et al.,
%    \emph{An Overview of Radiation-Induced Interface Traps in MOS Structures},
    Semicond. Sci. Technol.~4~(1989)~986, DOI:~10.1088/0268-1242/4/12/004.

 \bibitem{Reed:1987}
   M.L.~Reed
%    \emph{Chemistry of Si-SiO$_2$ interface-trap annealing},
    J. Appl. Phys.~63~(1988)~5776, DOI:~10.1063/1.340317.

 \bibitem{Poehlsen1:2012}
  T. Poehlsen et al.,
   \emph{Time dependence of charge losses at the Si-SiO$_2$ interface in $p^+n$-silicon strip sensors},
    to be published in Proceedings of PIXEL2012, submitted to Nucl. Instr. and Meth. A.

% ---- old ----
% \bibitem{Longoni:1990}
%  A.~Longoni, M.~Sampietro and L.~Strüder,
%   \emph{Instability of the behaviour of high resistivity silicon detectors due to the presence of oxide charges}, Nucl. Instr. and Meth. A~288~(1990)~35.

% \bibitem{Hartjes:2005}
%  F.G. Hartjes,
%   \emph{Moisture sensitivity of AC-coupled silicon strip sensors}, Nucl. Instr. and Meth. A~552~(2005)~168.

% \bibitem{Richter:1996}
%  R.H.~Richter et al.,
%   \emph{Strip detector design for ATLAS and HERA-B using two-dimensional device simulation},
%     Nucl. Instr. and Meth. A~377~(1996)~412.

% \bibitem{Hamamatsu}
%  \url{http://www.hamamatsu.com/}.

% \bibitem{Becker:Thesis}
%  J.~Becker,
%   \emph{Signal development in silicon sensors used for radiation detection}, PhD thesis, Universität Hamburg, DESY-THESIS-2010-33 (2010).

% --- References from previous paper --------------

%\bibitem{Kemmer:1980}  J.~Kemmer,  \emph{Fabrication of low noise silicon radiation detectors by the planar process},    Nucl. Instr. and Meth.~169~(1980)~499.

%\bibitem{Kemmer:1984}  J.~Kemmer,  \emph{Improvement of detector fabrication by the planar process},    Nucl. Instr. and Meth.~226~(1984)~89.

%\bibitem{Nicollian:1982}  E.H.~Nicollian and J.R.~Brews,  \emph{MOS (Metal Oxide Semiconductor) Physics and Technology},     New York, Wiley-Interscience, 1982.

%\bibitem{Richter:1996}  R.H.~Richter et al.,  \emph{Strip detector design for ATLAS and HERA-B using two-dimensional device simulation},     Nucl. Instr. and Meth. A~377~(1996)~412.

%\bibitem{Eremin:2003}  V.~Eremin et al.,  \emph{The charge collection in single side silicon microstrip detectors},
%    Nucl. Instr. and Meth. A~500~(2003)~121.

%\bibitem{Verbitskaya:2003}  E.~Verbitskaya et al.,  \emph{Effect of SiO2 Passivating Layer in Segmented Silicon Planar Detectors on the Detector Response},
% IEEE TRANSACTIONS ON NUCLEAR SCIENCE, VOL. 52, NO. 5, OCTOBER 2005.


%\bibitem{CIS} \url{http://www.cismst.org/}.

%\bibitem{Zhang:2011a} J.~Zhang et al., \emph{Study of radiation damage induced by 12~keV X-rays in MOS structures built on high resistivity n-type silicon}, Journal of Synchrotron Radiation, 19~(2012)~340, and arXiv 1107.5949.

%\bibitem{Zhang:2011b}  J.~Zhang et al.,   \emph{Study of X-ray Radiation Damage in Silicon Sensors},  JINST~6~C11013~(2011).

%\bibitem{Perrey:Thesis}   H.~Perrey,    \emph{Jets at Low Q2 at HERA and Radiation Damage Studies for Silicon Sensors for the XFEL},     PhD thesis, Universität Hamburg, DESY-THESIS-2011-021 (2011).

% \bibitem{Zhang:Thesis}   J.~Zhang,    \emph{X-ray Radiation Damage Studies and Design of a Silicon Pixel Sensor for Science at the XFEL}, PhD thesis, Universität Hamburg, in preparation.

% \bibitem{Kraner:1993}  H.W.~Kraner, Z.~Li and E.~Fretwurst,   \emph{The Use of the Signal Current Pulse Shape to Study the Internal Electric Field Profile and Trapping Effects in Neutron Damaged Silicon Detectors}, Nucl. Instr. and Meth. A~326~(1993)~350.

%\bibitem{Becker:2011}  J.~Becker, D.~Eckstein, R.~Klanner and G.~Steinbrück,   \emph{Measurements of charge carrier mobilities and drift velocity saturation in bulk silicon of   $\langle 111 \rangle $ and   $\langle 100 \rangle $ crystal orientation at high electric fields}, Solid State Electronics, 56~(2011)~104.


%\bibitem{Synopsys}
% Synopsys TCAD webpage:  \url{http://www.synopsys.com}.


%\bibitem{Poehlsen:Thesis}
%  T.~Pöhlsen,
%   \emph{Charge collection in irradiated silicon sensors}, PhD thesis, Universität Hamburg, in preparation.

\end{thebibliography}
\end{document}